\documentclass[final,twocolumn,3p,times]{elsarticle}
\usepackage{graphicx}
\usepackage{amssymb}
\usepackage{amsmath}
\usepackage{hyperref}
\usepackage{color}

\journal{Physica E}

\begin{document}

 \begin{frontmatter}

\title{Phase-separated symmetry-breaking   vortex-lattice  in a binary  Bose-Einstein condensate}

\author[ift]{Sadhan K. Adhikari\corref{author}}
\ead{sk.adhikari@unesp.br}
\cortext[author]{Corresponding author}


  \address[ift]{Instituto de F\'{\i}sica Te\'orica, Universidade Estadual
                     Paulista - UNESP,  01.140-070 S\~ao Paulo, S\~ao Paulo, Brazil}

\begin{abstract}

We study spontaneous-symmetry-breaking  circularly-asymmetric   phase separation of vortex lattices in a rapidly rotating harmonically-trapped
quasi-two-dimensional  (quasi-2D)  binary Bose-Einstein condensate (BEC) with repulsive inter- and intra-species interactions.  The phase separated vortex lattices of the components appear in different regions of space with no overlap between the vortices of the two components, which will permit an efficient experimental  observation of such vortices  and accurate study of the effect of
atomic interaction on such vortex  lattice. Such phase separation takes place when the intra-species interaction energies of the two components are equal or nearly equal with relatively strong inter-species repulsion.  When the  intra-species energies are equal, the two phase-separated vortex lattices have identical semicircular shapes with one being the parity conjugate of the other. When the  intra-species energies are nearly equal,    the phase separation is also complete but the vortex lattices have different shapes.  We demonstrate our claim with a numerical solution of the mean-field Gross-Pitaevskii equation for a rapidly rotating quasi-2D binary BEC.

\end{abstract}



\begin{keyword}

 Binary Bose-Einstein condensate,  Vortex-lattice formation, Gross-Pitaevskii equation, Spontaneous symmetry breaking

\end{keyword}

\end{frontmatter}

\section{Introduction}

 Soon after the observation  of  ultra-dilute and ultra-cold  trapped Bose-Einstein condensate (BEC) of alkali-metal atoms  in a laboratory \cite{becexpt,becexpt2}, rapidly rotating 
trapped condensates  were created and studied. A small number of vortices were created \cite{vors} for a small angular frequency of rotation $\Omega$.  Large vortex arrays 
were generated \cite{vorl} with the increase of $\Omega$. Similar to the vortices in  super-fluid $^4$He in a container, the vortices in trapped BEC also  
have quantized circulation: \cite{Sonin,fetter}
$\oint_{\cal C} {\bf v} .  d{\bf r}={2\pi\hbar l}/{m},$
where  ${\cal C}$ is a generic closed path,
${\bf v}({\bf r},t)$ is the super-fluid velocity field at the space point $\bf r$ and time $t$, $l$ is quantized integral  angular momentum of an atom in units of $\hbar$ in the trapped rotating BEC, and 
$m$ is the mass of an atom.  
As $\Omega$  increases,    it is energetically favorable to form a lattice of  vortices of unit circulation each ($l=1$) \cite{fetter}.
Consequently, a rapidly rotating trapped BEC generates  a large number of vortices of unit circulation  arranged usually  in a Abrikosov  triangular lattice \cite{vorl,abri}.   The ultra-dilute trapped BEC is formed  in the perturbative weak-coupling   mean-field limit. This allows      to study the formation of  vortices in such a   BEC  by the mean-field   Gross-Pitaevskii (GP) equation.

Vortex lattice formation in a trapped rotating binary BEC   with a large number of vortices has also been observed \cite{Schweikhard,Hall} and studied theoretically \cite{wang,thlatsep}.
There has also been study of vortex-lattice formation in a BEC along the weak-coupling to unitarity crossover \cite{sci}  and in a rotating box trap \cite{jpcm}.
 The study of vortex lattices in a binary or a multi-component spinor BEC is interesting because the interplay between intra-species and inter-species interactions may lead to the formation of  square    \cite{Schweikhard,kumar}, stripe  and honeycomb 
 \cite{honey} vortex lattice, other than the standard Abrikosov triangular lattice \cite{abri}.    In addition, there could be
the formation of coreless vortices \cite{coreless}, 
  vortices of fractional charge \cite{frac,Cipriani},   and phase-separated vortex lattices in multi-component non-spinor \cite{cns}
spinor  \cite{thlatsep} and dipolar  \cite{kumar}   BECs. The difficulty of 
experimental study of overlapping vortices in  different components of  a multi-component or a binary BEC is monumental and despite great interest in the study of vortex lattices in a binary BEC 
\cite{Mueller,Barnett,Wei,Kuopanportti,Kasashet, Kasarev2},
 this has highly limited such experimental   studies \cite{Schweikhard,Hall}. Hence for experimental study  it will be highly desirable  to have phase-separated vortex lattices in a binary BEC, where the vortices and component density  of one component do not overlap with those of the other.

In a repulsive homogeneous BEC,  phase separation takes place for  \cite{phse}
\begin{equation} \label{eq1}
\frac{g_1g_2}{g_{12}^2}<1, 
\end{equation}
where $g_1$ and $g_2$ are intra-species repulsion strengths for components 1 and 2, respectively, and $g_{12}$ inter-species repulsion strength.  This useful condition, although not rigorously valid in a trapped quasi-two-dimensional (quasi-2D)  BEC, may
provide an approximate guideline for phase separation. 
It is well-known  that a phase separated vortex lattice in a rotating binary BEC can be generated  by manipulating the parameters $g_1,g_2,$ and $g_{12}$ \cite{thlatsep,phse}.
In this paper we identify the parameter domains leading to completely phase-separated vortex lattices in a rotating quasi-2D binary BEC so that the vortices of one component does not have any overlap with the matter density of the other component. 
We  find that  if $g_1$ and $g_2$ are equal and $g_{12}$ is much larger than $g_1$, then the phase separated components have the form of completely non-overlapping semi-circles lying opposite to each other  spontaneously breaking the circular symmetry of the underlying Hamiltonian.  In this case the vortex-lattice formation on the two component semi-circles has triangular symmetry with one component completely avoiding the other.   
 If $g_1$ and $g_2$ are largely different and satisfy condition (\ref{eq1}),  one of the components lie on a  circle at the center of the trap with the second component lying on a concentric circular anel outside the first component maintaining the circular symmetry for a non-rotating binary BEC. In the case of a rotating binary BEC, although there will be phase-separated vortex lattices, 
the component densities will  have irregular shape breaking the circular symmetry.  
If    $g_1$ and $g_2$ are nearly equal and $g_{12}$ much larger, 
the fully phase-separated components break  circular symmetry and have different shapes. If such a phase-separated spontaneous-symmetry breaking  binary BEC  
is subject to a rapid rotation, the generated vortex lattices maintain the same shape as the non-rotating binary BEC  breaking the circular symmetry spontaneously.    
For $g_1=g_2$, the phase-separated vortex lattices are  dynamically stable. 
  In this paper we will study numerically the generation of vortex lattices in these  cases using the mean-field GP equation.

In Sec. II  the mean-field model for a   rapidly rotating binary BEC is presented.  Under a tight trap in the transverse direction a quasi-2D version of the model is also given, which we use in the present study. 
The results of numerical calculation are shown in Sec. III.  
Finally, in Sec. IV we present a brief summary of our findings.

\section{Mean-field model for a rapidly rotating binary BEC}

We consider a binary rotating BEC  interacting via  inter- and intra-species 
interactions.  
The angular frequencies for the axially-symmetric harmonic trap  
along $x$, $y$ and $z$ directions are taken as 
$\omega_x=\omega_y=\omega$ and 
$\omega_z=\lambda\omega$,  respectively.    Now it is possible to make a binary BEC of two hyper-fine states of the same atomic species, such as $^{39}$K and  $^{87}$Rb  atoms.
   In such cases   the masses of the two species are equal. Thus in this theoretical study we will take the masses of two species to be equal.  

The study of a rapidly rotating binary BEC is conveniently performed in the rotating frame, where the generated vortex  
lattice is a stationary state \cite{fetter}, which can be obtained by the imaginary-time propagation method \cite{imag}. 
Such a dynamical equation in the rotating frame can be written if we note that the Hamiltonian in the rotating frame is given by  $H = H_0-\Omega l_z$, where $H_0$ is that in the laboratory frame, $\Omega $  is the angular frequency of rotation,  $l_z$ is the $z$ component of angular momentum given by $l_z= i\hbar (y\partial/\partial  x - x \partial/\partial y )$ \cite{ll1960}.
However, if the rotational frequency $\Omega$ is increased beyond the trapping frequency $\omega$, the rotating bosonic gas makes a quantum phase transition to a non-super-fluid state, where the validity of a mean-field description of the rotating bosonic gas is questionable \cite{fetter}.  Hence it is appropriate to limit this study to $\Omega< \omega$.  
With the inclusion of the extra rotational energy  $-\Omega l_z$ in the Hamiltonian,   the coupled GP
equations for the binary  BEC in the rotating frame for $\Omega <\omega $  can be written as \cite{quartic}
\begin{align} \label{eq1x}
{\mbox i} \hbar \frac{\partial \phi_1({\bf r},t)}{\partial t} &=
{\Big [}  -\frac{\hbar^2}{2m}\nabla^2 -\Omega l_z  
+ \frac{1}{2}m \omega^2 
(\rho^2 +\lambda^2{z}^2 )
\nonumber \\  &  
+\frac{4\pi \hbar^2}{m} \Big\{{a}_1 N_1 \vert \phi_1({\bf r},t)\vert^2
\nonumber
\\  &  
+ {a}_{12} N_2 \vert \phi_2({\bf r},t)|^2\Big\}
{\Big ]}  \phi_1({\bf r},t),
\\  
\label{eq2}
{\mbox i} \hbar \frac{\partial \phi_2({\bf r},t)}{\partial t} &=
{\Big [}  -\frac{\hbar^2}{2m}\nabla^2 -\Omega l_z + \frac{1}{2}m \omega^2 
(\rho^2+\lambda^2{z}^2 )
\nonumber\\ &
+ \frac{4\pi \hbar^2}{m}\Big\{ {a}_2 N_2 \vert \phi_2({\bf r},t) \vert^2
\nonumber\\ &
+ {a}_{12} N_1 \vert \phi_1({\bf r},t) \vert^2\Big\}
\Big] 
 \phi_2({\bf r},t),
\end{align}
where   the two species of atoms of mass $m$ each are denoted $i=1,2$, $\phi_i({\bf r},t)$ are the order parameters of the two components, 
$N_i$ is the number of atoms in species 
$i$,  $\quad {\mbox i}=\sqrt{-1}$, ${\bf r}= \{x,y,z\},$  $ {\pmb \rho}=\{ x,y\}$, $\rho^2=x^2+y^2$, $a_i$ is the intra-species scattering length of species $i$,  $a_{12}$ is the inter-species scattering length. The functions $\phi_i$ are normalized as $\int d{\bf r}|\phi_i({\bf r},t)|^2 =1.$

The following   dimensionless form of Eqs. (\ref{eq1x}) and (\ref{eq2})  can be obtained  by  the  transformation of variables: ${\bf r}' = {\bf r}/l_0, l_0\equiv \sqrt{\hbar/m\omega}$, $t'=t\omega,  \phi_i'=   \phi_i l_0^{3/2},  \Omega'=\Omega/\omega, l_z '= l_z/\hbar$ etc.:   
\begin{align}
{\mbox i} \frac{\partial \phi_1({\bf r},t)}{\partial t}=& \,
{\Big [}  -\frac{\nabla^2}{2 }
+ \frac{1 }{2} (\rho^2 +\lambda^2 z^2 ) -\Omega l_z
+ 4\pi N_1a_1 \vert \phi_1 \vert^2 
\nonumber\\ &
+ 4\pi a_{12}N_2\vert \phi_2 \vert^2
{\Big ]}  \phi_1({\bf r},t),
\label{eq3}\\
{\mbox i} \frac{\partial \phi_2({\bf r},t)}{\partial t}=& \,{\Big [}  
- \frac{\nabla^2}{2}+ \frac{1 }{2} (\rho^2+\lambda^2 z^2 ) -\Omega l_z
+ 4\pi N_2 a_2 \vert \phi_2 \vert^2 
\nonumber\\ &
+ 4\pi a_{12} N_1 \vert \phi_1 \vert^2 
{\Big ]}  \phi_2({\bf r},t),
\label{eq4}
\end{align} 
where for simplicity we have dropped the prime from the transformed variables.

For a quasi-2D binary BEC in the $x-y$ plane under a strong trap along the $z$ 
direction ($\lambda \gg 1$), the essential vortex dynamics will be confined to  the $x-y$ plane with the $z$ dependence playing a passive role.   The wave functions  can then be written as 
$\phi_i({\bf r},t)= \psi_i({\pmb \rho},t)\Phi(z)$, where the function $ \psi_i({\pmb \rho},t)$ carries the essential vortex dynamics and $\Phi(z)$ is a normalizable Gaussian function. In this case the 
$z$ dependence can be integrated out \cite{luca} and we have the following 2D equations
\begin{align}
{\mbox i} \frac{\partial \psi_1({\pmb \rho},t)}{\partial t}=& \,
{\Big [}  -\frac{\nabla^2}{2 }
+ \frac{1 }{2} \rho^2  -\Omega l_z
+ g_1 \vert \psi_1 \vert^2
\nonumber\\ &
 + g_{12} \vert \psi_2 \vert^2
{\Big ]}  \psi_1({\pmb \rho},t),
\label{eq5}\\
{\mbox i} \frac{\partial \psi_2({\pmb \rho},t)}{\partial t}=& \, {\Big [}  
- \frac{\nabla^2}{2}+ \frac{1 }{2} \rho^2 -\Omega l_z 
+ g_2 \vert \psi_2 \vert^2 
\nonumber\\ &
+ g_{21} \vert \psi_1 \vert^2 
{\Big ]}  \psi_2({\pmb \rho},t),
\label{eq6}
\end{align}
where
$g_1=4\pi a_1 N_1\sqrt{\lambda/2\pi},$
$g_2= 4\pi a_2 N_2\sqrt{\lambda/2\pi}  ,$
$g_{12}={4\pi } a_{12} N_2 \sqrt{\lambda/2\pi} ,$
$g_{21}={4\pi } a_{12} N_1  \sqrt{\lambda/2\pi}.$ In this study we will take $N_1=N_2$ which will make $g_{12}=g_{21}$ maintaining the possibility $g_1\ne g_2$  and consider $\Omega<1$ \cite{fetter}.

The binary GP equations (\ref{eq5}) and (\ref{eq6}) can also be obtained using a variational procedure:
\begin{equation}
 {\mbox i}\frac{\partial \psi_i({\pmb \rho},t)}{\partial t}=\frac{\delta E}{\delta \psi_i^*({\pmb \rho},t)}
\end{equation}
with the following energy functional in the rotating frame:
\begin{align}\label{en}
E[\psi] = \int d {\pmb \rho} \Big[\sum_ i \frac{1}{2}   \biggr( |\nabla \psi_i|^2 + {\rho^2} |\psi_i|^2+ g_i|\psi_i|^4
\nonumber \\
- 2\psi_i^*l_z \Omega \psi_i \biggr) 
+ g_{12}|\psi_1|^2|\psi_2|^2\Big] .
\end{align}
We note that the rotational  energy {  $ -\int d {\pmb \rho}\psi_i^*l_z \Omega \psi_i $  is negative.} Hence the energy $E$ in the rotating frame  will decrease with the increase of angular frequency of rotation. All other contributions to energy (\ref{en}) are positive. Hence in the perturbative limit of small $\Omega$,  the total energy will be positive with  the energy decreasing linearly with $\Omega$. { For very  large $\Omega$ $(\Omega < 1,$  $ \Omega \to 1)$,} the contribution of the rotational energy  will be proportional to $\Omega^2$  \cite{fetter2} and the total energy will decrease quadratically with $\Omega$, {  viz. Fig. \ref{fig5}(b) below. }

\begin{figure}[!t]

\begin{center}
\includegraphics[trim = 0cm 0.0cm 0cm 0mm, clip,width=\linewidth]{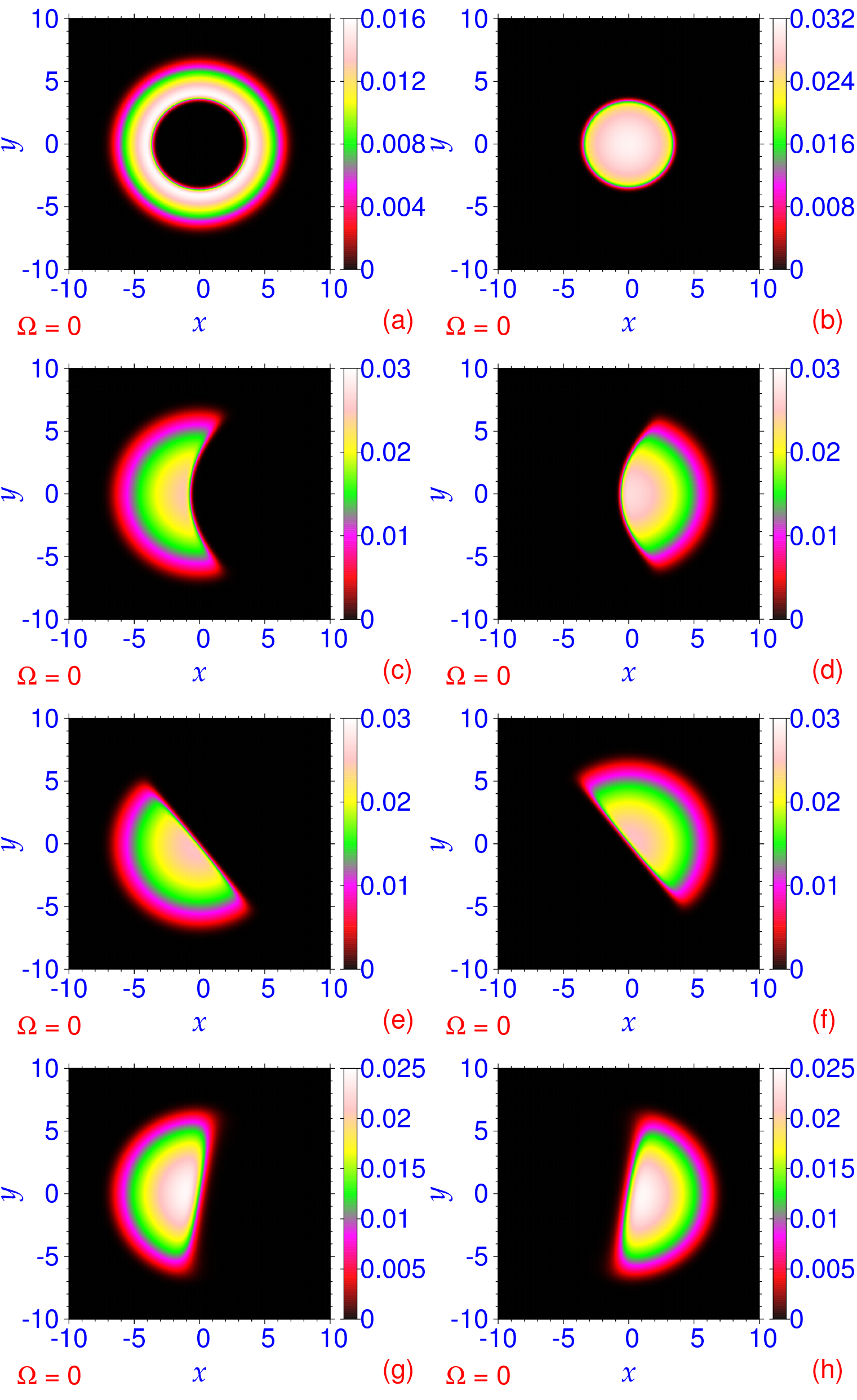} 

\caption{   Phase separation     in a non-rotating ($\Omega =0$) binary BEC from a contour plot of 2D densities ($|\psi_i|^2$): (a) first and (b) second components for $g_{12}=2000, g_1= 1000, g_2=700$, (c) first and (d) second components for 
$g_{12}=2000, g_1=1000, g_2=900 $ (e) first and (f) second components for $g_{12}=2000$,  $g_1=g_2=1000 $, (g) first and (h) second components for $g_{12}=1100, g_1=g_2= 1000$.
 All quantities plotted in this and following figures are dimensionless.
}
\label{fig1}
\end{center}

\end{figure}

\section{Numerical Results}

\begin{figure}[!t]

\begin{center}
\includegraphics[trim = 0cm 0.64cm 0cm 0mm, clip,width=\linewidth]{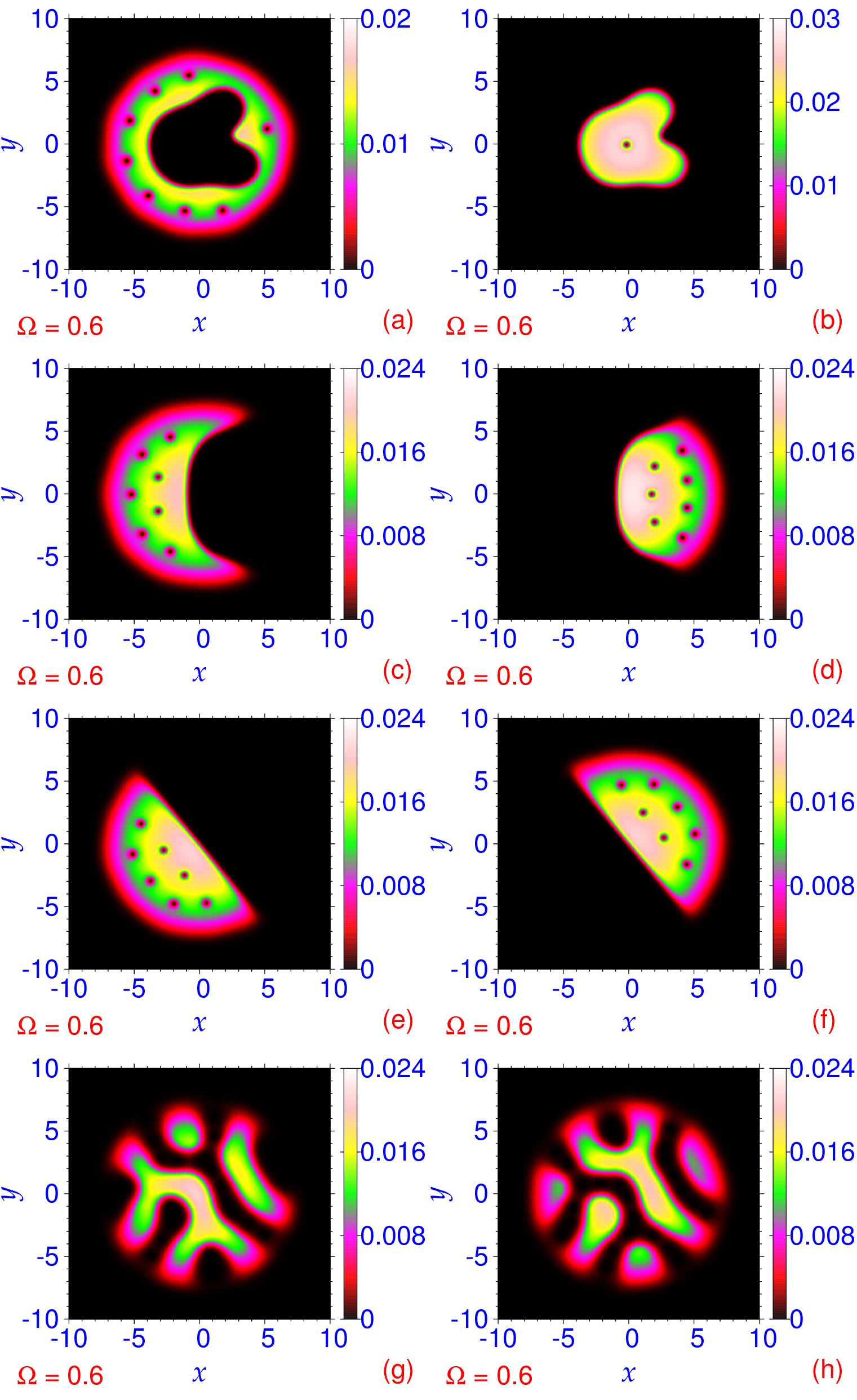} 
 
\caption{   Phase-separated vortex lattices in a rapidly rotating binary BEC with $\Omega=0.6$  from a contour plot of 2D densities ($|\psi_i|^2$): (a) first and (b) second components for $g_1=1000, g_2=700, g_{12}=2000$, (c) first and (d) second components for  $g_1=1000, g_2=900, g_{12}=2000$,   (e) first and (f) second components for $g_1=g_2=1000, g_{12}=2000$,  (g) first and (h) second components for $g_1=g_2=1000,  g_{12}=1100$, corresponding to the non-rotating BECs shown in Fig. \ref{fig1}(a)-(h), respectively.  }
\label{fig2}
\end{center}

\end{figure}

The quasi-2D binary mean-field equations
 (\ref{eq5}) and (\ref{eq6}) cannot be solved analytically and different numerical methods, such as the split time-step Crank-Nicolson method \cite{imag,CPC} or the pseudo-spectral method \cite{PS}, can be employed  for their solution.
Apart from the basic 
C and FORTRAN programs for solving the GP equation \cite{imag}, their open multi-processing (OMP)
versions \cite{CPC} are also available
  and one should use the appropriate one.  The OMP versions reduce the execution time significantly in a multi-core multi-processing computer. 
These OMP  programs have recently been adapted to simulate the vortex lattice in a rapidly rotating BEC \cite{cpckk} and we use these in this study.  Here we solve  Eqs. (\ref{eq5}) and (\ref{eq6}) by the split time-step
Crank-Nicolson discretization scheme using a space step of 0.05
and a time step of 0.0002 for imaginary-time simulation and 0.0001 for real-time simulation.  
The imaginary-time simulation is performed with a localized initial state modulated by a random phase at each space grid point as in Refs. \cite{jpcm,cpckk}.  The random phase modulation allows an efficient generation of vortex lattice independent of the algebraic form of the initial state.   The real-time simulation is performed with the converged imaginary-time wave function as the initial state.

 In this paper, without considering a specific atom, we will present the results in dimensionless units for different sets of  parameters: 
$\Omega, g_1, g_2, g_{12} (=g_{21})$. 
In the phenomenology of a specific atom, the parameters   $g_1, g_2, g_{12}$ can be varied experimentally through a variation of the underlying intra- and inter-species scattering 
lengths 
 by the Feshbach resonance technique \cite{fesh}.

\begin{figure}[!t]

\begin{center}
\includegraphics[trim = 0cm 0.cm 0cm 0mm, clip,width=\linewidth]{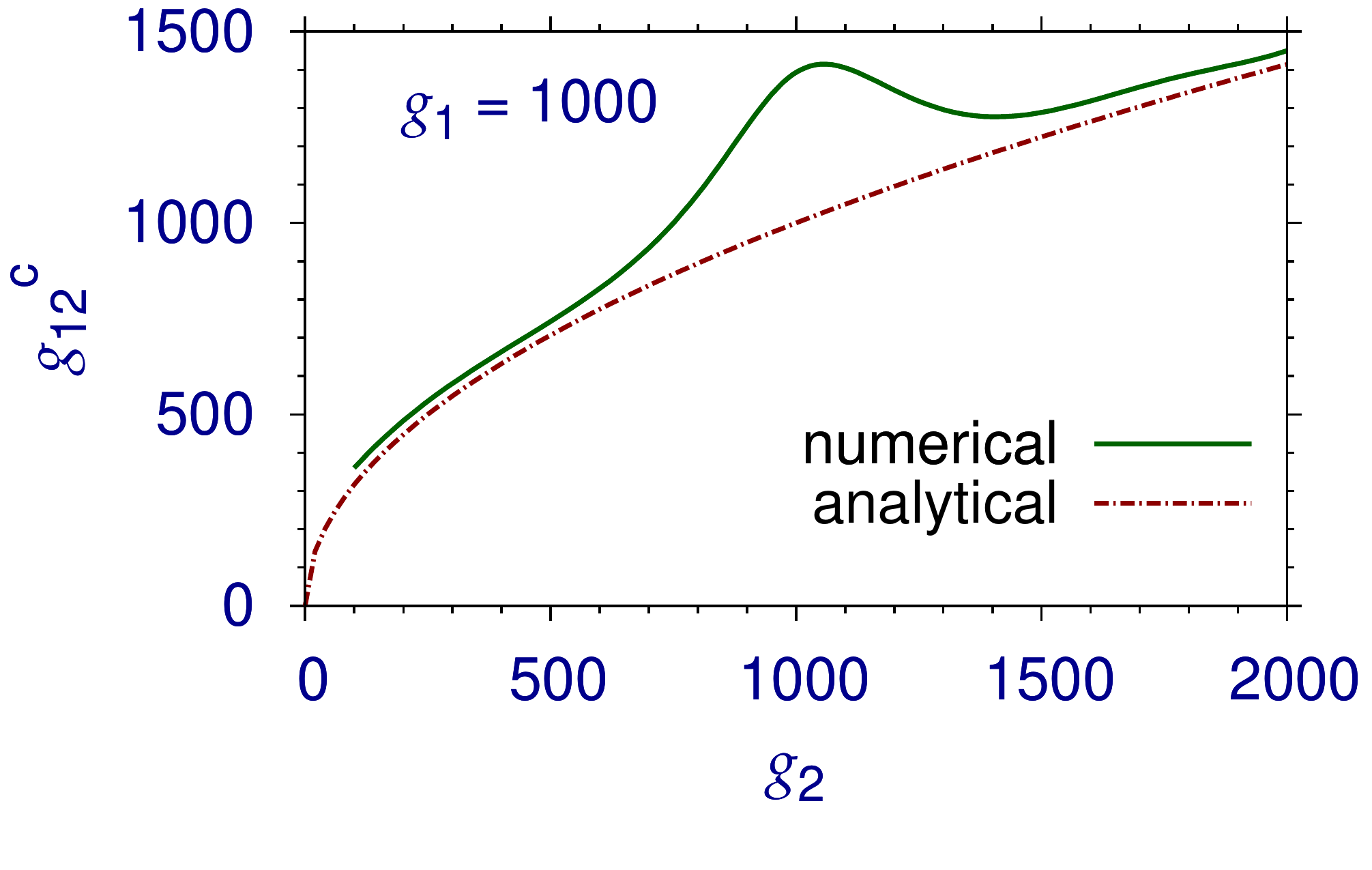}

\caption{ The numerical $g_{12}^c$ versus $g_2$ for $g_1=1000$  and its analytical estimate  $g_{12}^c=\sqrt{g_1g_2}$ from Eq. (\ref{eq1}).   The numerical estimate is found to be independent of the angular frequency of rotation $\Omega$. }
\label{fig3}
\end{center}

\end{figure}

First we   demonstrate a phase separation  in a fully repulsive ($g_i,g_{12}>0$) non-rotating binary BEC ($\Omega=0$).
We find that a phase separation generally follows the condition (\ref{eq1}) for $g_1 \approx g_2$; so that $g_{12}>g_1,g_2$. 
For our purpose,  we consider $g_1=1000, g_{12}=2000$ and $g_2=$ (i) 700, (ii) 900, (iii) 1000, and  (iv) $g_1=g_2=1000$ with $g_{12}=1100.$ The result of phase separation is shown in Figs. \ref{fig1}(a)-(h), where we plot  density of the two components. 
 In case (i) $g_1 =1000$ and $g_2=700$ are quite different maintaining $ g_{12}=2000$ much larger than $g_1$ and $g_2$ and phase separation maintains circular symmetry  as shown in Figs.  \ref{fig1}(a)-(b). As $g_2=900$ approaches $g_1=1000$ in case (ii), the phase separation spontaneously breaks the circular symmetry of the Hamiltonian as   shown in Figs.  \ref{fig1}(c)-(d).  In case (iii), $g_1=g_2=1000$ and the phase separation breaks the circular symmetry, viz.  Figs.  \ref{fig1}(e)-(f),  with the density of component 1 being the parity conjugate of the density of component 2. Finally, in case (iv) $g_{12}=1100$ approaches $g_1=g_2=1000$  consistent with condition   (\ref{eq1}), maintaining a parity symmetric phase separation. 
In this paper, we will be interested in robust spontaneous-symmetry breaking phase separated vortex lattices without overlap between component densities, which is of great experimental interest. Hence cases (ii), (iii), and (iv) seem to be attractive candidates  and we next study the generation of vortex lattices in the cases displayed in Fig. \ref{fig1}.

In Figs. \ref{fig2}(a)-(h) we plot the component densities of a rotating binary BEC with angular frequency of rotation
 $\Omega =0.6$ with same interaction parameters $g_1,g_2, g_{12}$ as the non-rotating BECs considered in Figs.  \ref{fig1}(a)-(h). The case (i) illustrated in Figs.  \ref{fig2}(a)-(b)  breaks the circular symmetry generating an irregular shape of the components  and hence an irregular arrangement of vortices and we will not study this case in detail.   In the non-rotating BEC with same parameters, the component densities maintain circular symmetry, viz.  Figs.  \ref{fig1}(a)-(b). In 
case (ii) displayed in Figs.  \ref{fig2}(c)-(d) the rotating BEC has density profile quite similar to the non-rotating   
BEC of Figs.  \ref{fig1}(c)-(d). The same is true in  Figs.  \ref{fig2}(e)-(f) when  compared with  Figs.  \ref{fig1}(e)-(f).
Nevertheless, the arrangement of vortices in Figs. \ref{fig2}(c)-(f)  is very ordered with a definite (triangular)  symmetry.
In Figs. \ref{fig2}(g)-(h), quite surprisingly, the phase separation found in the non-rotating BEC of Figs. \ref{fig1}(g)-(h) has disappeared generating   two component BECs occupying the whole region of space. Moreover, no visible vortices are found in Figs. \ref{fig2}(g)-(h). This type of density distribution is called vortex sheet structure  which has been found before in binary BECs \cite{honey,Kasashet}.  This case is not of interest of the present study. In the following we will study vortex lattice generation in cases (ii) and (iii) in some detail  with a triangular arrangement of vortices in the components and with complete phase separation of the density of the two components.

\begin{figure}[!t]

\begin{center}
\includegraphics[trim = 0cm 0.cm 0cm 0mm, clip,width=\linewidth]{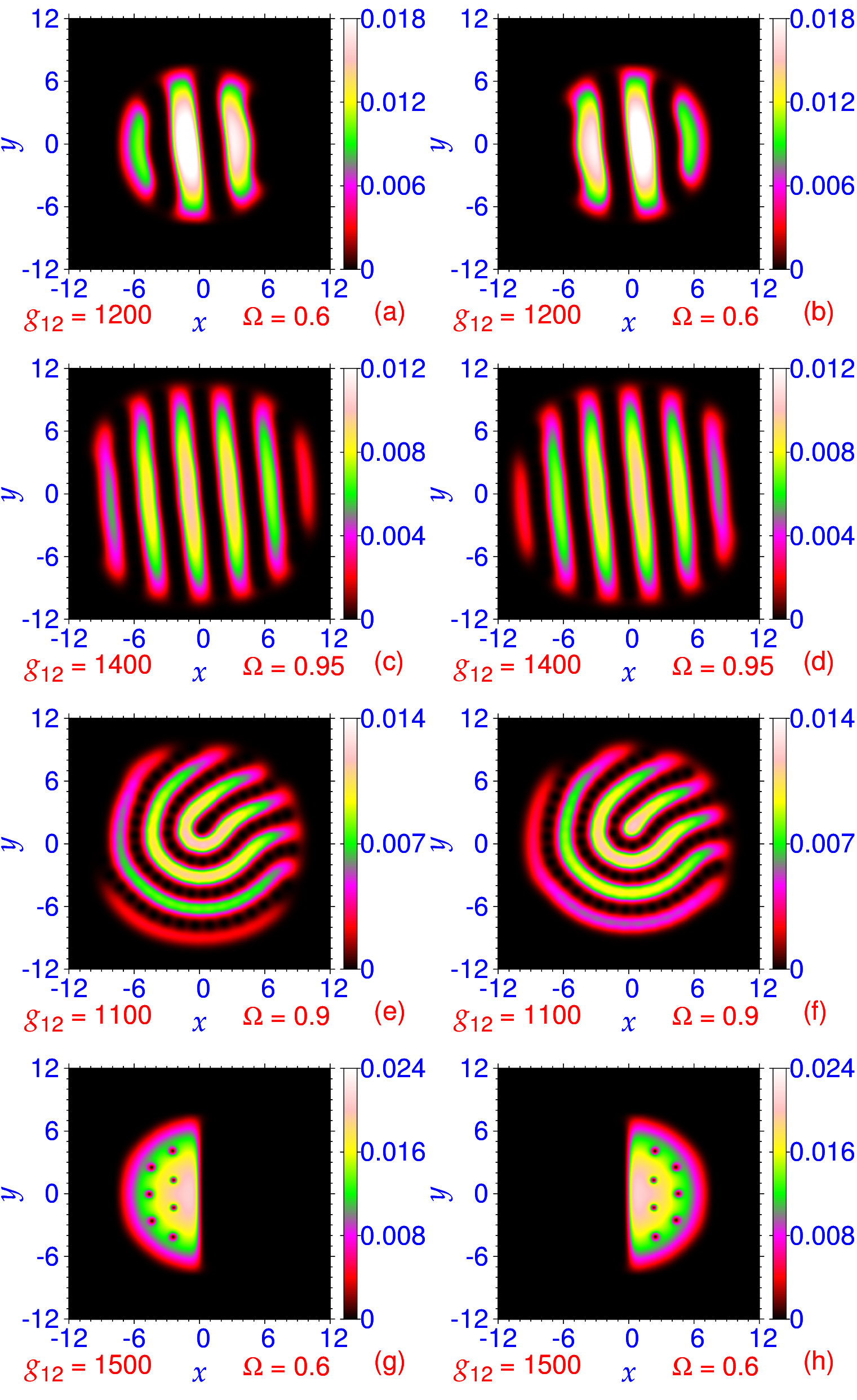}

\caption{ Contourplot of 2D  density of a rapidly rotating binary BEC for $g_1=g_2 = 1000$. (a) first and (b) second components for $g_{12}-1200, \Omega =0.6$;  (c) first and (d) second components for $g_{12}-1400, \Omega =0.95$;  (e) first and (f) second components for $g_{12}-1100, \Omega =0.9$; (g) first and (h) second components for $g_{12}-1500, \Omega =0.6$.}
\label{fig4}
\end{center}

\end{figure}

\begin{figure}[!t]

\begin{center}
\includegraphics[trim = 0cm 0.cm 0cm 0mm, clip,width=\linewidth]{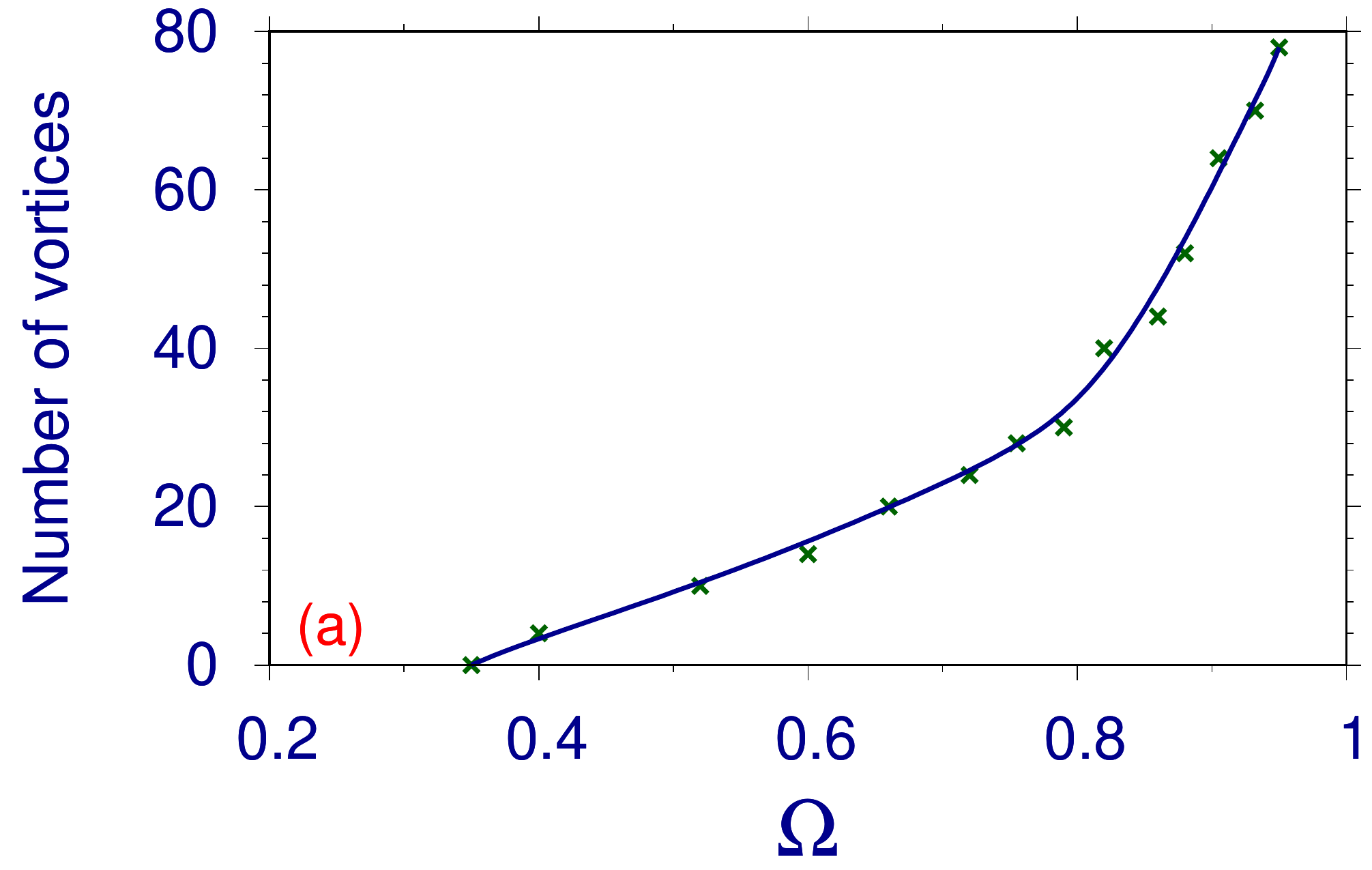}
\includegraphics[trim = 0cm 0.cm 0cm 0mm, clip,width=\linewidth]{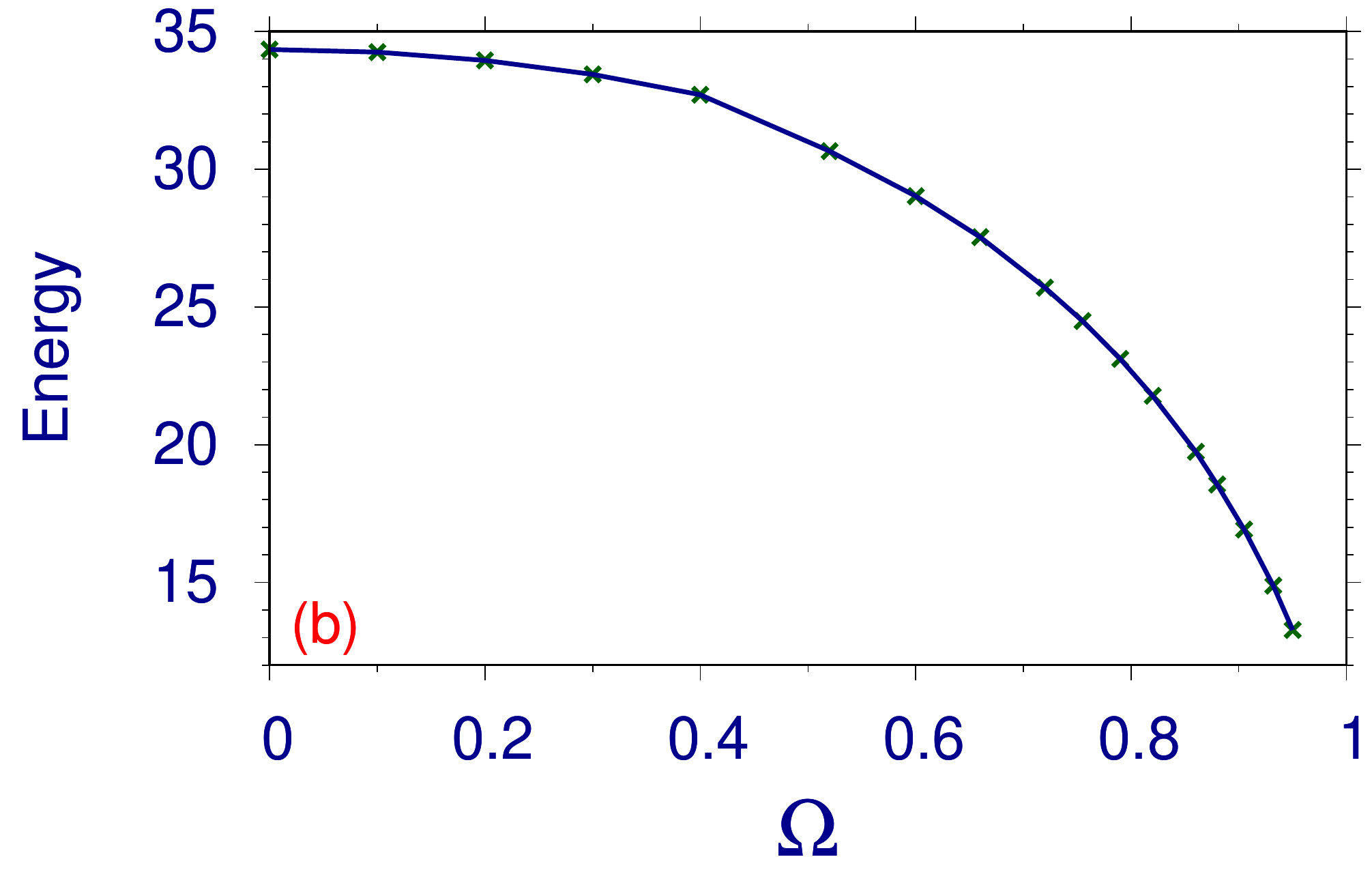} 
 
\caption{  (a) Number of vortices and (b) energy in the rotating frame (\ref{en}) for a
rapidly rotating quasi-2D binary BEC   with
$g_1=g_2 = 1000, g_{12}=2000$ versus angular frequency of rotation $\Omega$. The crosses are the actual points
obtained numerically whereas the   lines are shown to guide the eye.  
}
\label{fig5}
\end{center}

\end{figure}

\begin{figure}[!t]

\begin{center}
\includegraphics[trim = 0cm 0.64cm 0cm 0mm, clip,width=\linewidth]{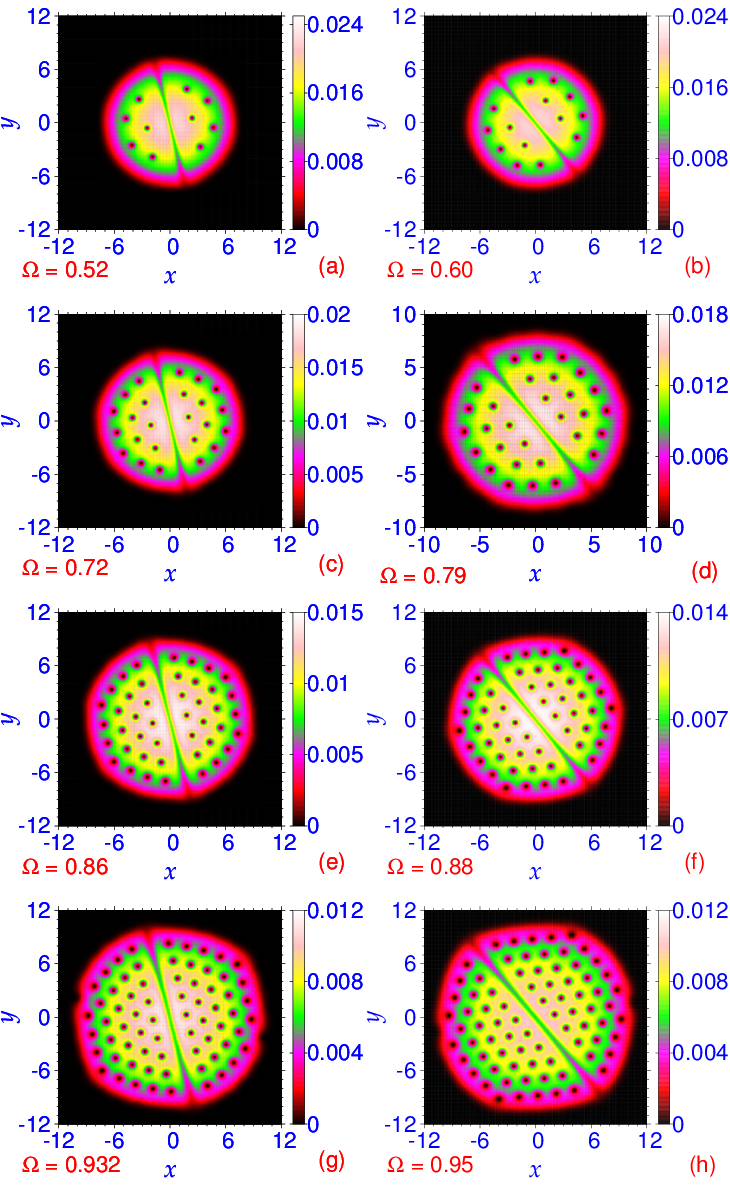} 

\caption{ First and  second components of a rotating binary BEC with angular frequency of rotation  $\Omega=$  (a)  0.52,
(b) 0.6, (c) 0.72, (d)  0.79, (e) 0.86, (f) 0.88, (g) 0.932, and (h) 0.95 with    the respective number of vortices  10, 14, 24, 30, 44, 52, 70, 80.
  Other parameters are 
 $g_1=g_2=1000, g_{12}=g_{21}=2000$.
}
\label{fig6}
\end{center}

\end{figure}

 First we study how the number of vortices and energy evolve with the increase of angular frequency of rotation $\Omega$.
For this purpose we consider the symmetric case $g_1=g_2 =1000$ with $g_{12}=  2000$.  We consider a large $g_{12}$ as for a large $g_{12}$ the phase separation is very stable  leading to a ground state with phase-separated vortex lattice.  For 
$g_{12}\gtrapprox g_1=g_2$, there is phase separation, but the ground state has a phase-separated sheet structure \cite{honey,Kasashet}, viz. Fig. \ref{fig2}(g)-(h). The fully phase-separated states  in this case with $g_1=g_2=1000, g_{12}=1100$, similar to those in  Figs. \ref{fig2}(e)-(f),  are  excited states of higher energy. 
 {  As $g_{12}$ increases, the difference of energy between these two types of  states reduce and beyond  a certain critical $g_{12}^c$ 
the fully phase-separated states become the ground state which can be studied experimentally.
In Fig. \ref{fig3} we plot the numerically computed   $g_{12}^c $  versus $g_2$ for a fixed $g_1 =1000$  and $\Omega =0.6$ and compare with the analytical result $g_{12}^c \equiv \sqrt{g_1g_2}= \sqrt{1000 g_2}$,
signaling a phase separation of a uniform (non-rotating) binary mixture, obtained from Eq. 
(\ref{eq1}). 
The numerical  critical value $g_{12}^c$  is found to be independent of  $\Omega$ $(<1)$. 
For values of $g_{12}$ larger than the numerical   $g_{12}^c$, phase separated vortex lattice is obtained. From Fig. \ref{fig3} we find that the formation of phase separated vortex lattice is most problematic for $g_1=g_2$ and we study the formation of phase separated vortex lattice  for this case in some details. 
 For values of   $g_{12}$ between the numerical and analytical estimates of  $g_{12}^c$, either    straight    stripe structure, viz. Fig. \ref{fig4}(a)-(d), or sheet or bent  stripe   structure, viz. Fig. \ref{fig2}(g)-(h) and   Fig. \ref{fig4}(e)-(f),   is formed for $g_1=g_2=1000$ \cite{Kasashet}. Only for   $g_{12}$
larger than its numerically obtained critical value, phase separated vortex lattice is formed as shown in   Fig. \ref{fig4}(g)-(h).  We find that the size and the number of stripes increase as the angular frequency of rotation is increased, viz. Fig. \ref{fig4}(a)-(d). The straight stripes   become bent  for smaller $g_{12}$, viz. Fig. \ref{fig4}(e)-(f), which change to sheet structure for small angular frequency $\Omega$, viz. Fig. \ref{fig2}(g)-(h). }

 We study the generation of vortex lattice for 
   $g_1=g_2 =1000$ with $g_{12}=  2000$. The  fully phase-separated states are the ground  states for all $\Omega$. In Figs. \ref{fig5}(a)-(b) we plot the number of vortices and energy in the rotating frame (\ref{en}) versus $\Omega$. As expected, the number of vortices increase with $\Omega $ and energy decreases as $\Omega$ is increased. The energy decreases as the contribution of the rotational energy  $-\Omega l_z$ in the expression for energy is negative. For small $\Omega$, in the perturbative limit, this contribution is linearly proportional to $-\Omega$. But as $\Omega$ increases, the number of vortices increase rapidly and in that non-perturbative domain the rotational energy behaves as $\Omega^2$ \cite{fetter}.  In Fig. \ref{fig5}(b), this transition from linear to quadratic dependence on $\Omega $ is seen as $\Omega$ increases.

\begin{figure}[!t]

\begin{center}
\includegraphics[trim = 0cm 0.64cm 0cm 0mm, clip,width=\linewidth]{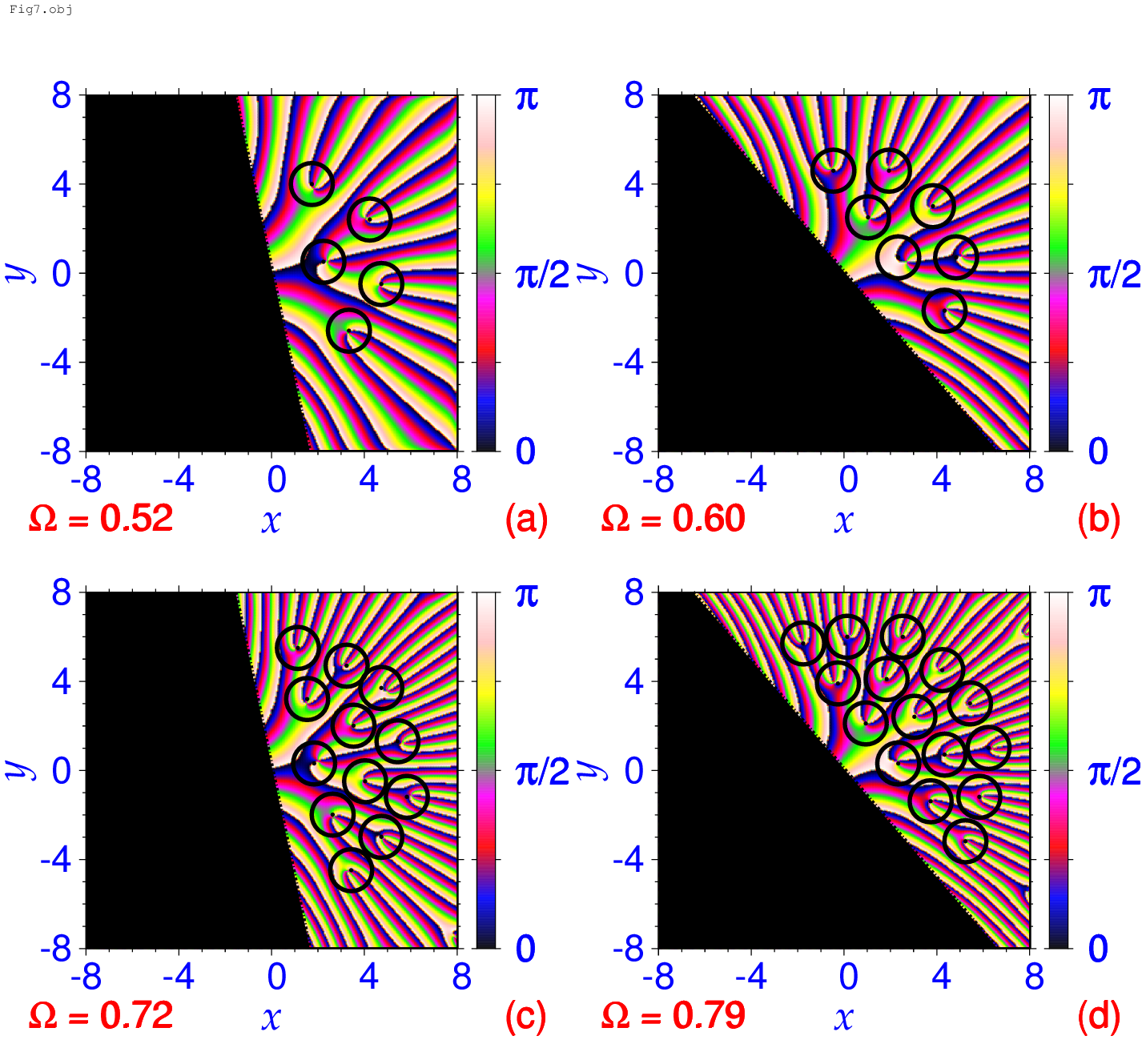}

\caption{ (a)-(d) A contour plot of the   phase of the second component of the wave function
on the consensate surface corresponding to the vortex lattice shown in Figs. \ref{fig6}(a)-(d), respectively.  The position of the vortices are marked by a closed circular contour.
}
\label{fig7}
\end{center}

\end{figure}

The generation of vortex lattice in the symmetric case $g_1=g_2$ reveals very interesting features.  The study is performed with the parameters $g_1=g_2=1000, g_{12}=2000$. Robust triangular vortex lattice is generated in this case on two completely separated semicircular components as shown in Figs. \ref{fig2}(e)-(f). The  generated vortex lattices for $\Omega =0.52, 0.60, 0.72, 0.79, 0.86, 0.88, 0.932,$ and  0.95  are shown in Figs. \ref{fig6}(a)-(h), respectively. As the phase separation is complete in this case we have displayed the vortex lattices of the two components of the binary BEC in the same plot. The separation between the two components can be clearly seen in this figure in a domain of low density between the two components. Comparing Figs. \ref{fig2}(e)-(f) on the one hand and Fig. \ref{fig6}(b) on the other, we find that  Fig. \ref{fig6}(b) provides a more concise illustration of the binary vortex lattice.  For the sake of convergence of the numerical scheme,  we used a larger domain of space in the numerical calculation  than the space domain  shown  in the plots of Fig. \ref{fig6}.   The maximum density of the condensate in these plots reduce with the increase of $\Omega$ as can be seen in this figure. With the increase of $\Omega$, due to increased centrifugal repulsion, the binary condensate  occupies a larger space domain thus reducing the maximum density. Another remarkable feature of the vortex lattice generation in Fig. \ref{fig6}  is the robust accurate triangular lattice formation in plots (e)-(h).

\begin{figure}[!t]

\begin{center}
\includegraphics[trim = 0cm 0.64cm 0cm 0mm, clip,width=\linewidth]{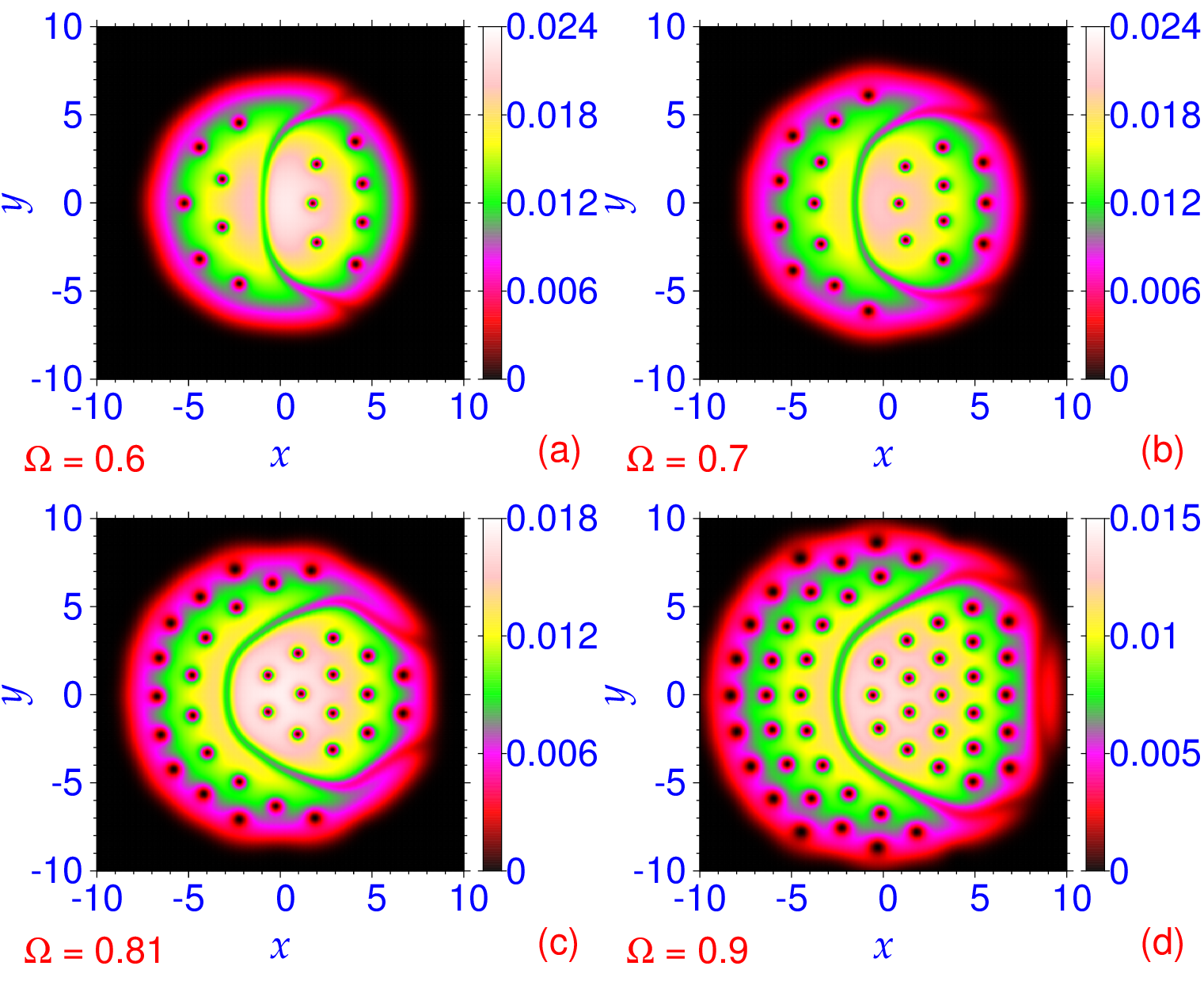}

\caption{   First and  second components of a rotating binary BEC with angular frequency of rotation  $\Omega=$  (a)  0.6,
(b) 0.7, (c) 0.81, and (d)  0.9, and $g_1=1000, g_2=900, g_{12}=g_{21}=2000$ with    the respective number of vortices   14 (7+7), 21 (11+10), 33 (19+14), 54 (31+23).
The vortex lattice of the first component with larger interaction strength ($g_1=1000$) than that of the second component ($g_2=900$)  in on the left side of each plot and that of the second component is on the right side.  
}
\label{fig8}
\end{center}

\end{figure}

{ 
To demonstrate that all spots in density of Fig. \ref{fig6} correspond to vortices, we display the phase  of the wave function $  \arctan [\Im(\psi_i)/\Re(\psi_i)]$
 on the condensate surface  in Figs. \ref{fig7}(a)-(d)  corresponding to the vortex 
lattice of Figs.  \ref{fig6}(a)-(d), where $\Im $ and $\Re$ are imaginary and real parts. Quantum vortices in a rotating super-fluid have  unit  circulation ($l=1$) each. 
   In a closed path around a  vortex of unit   angular momentum,   the phase changes by $2\pi$.  In Figs. \ref{fig7} the phase over the second component  is shown and the position of the vortices are indicated by a closed contour. A careful survey of these phases reveals that  the accumulated phase of the wave function
around the clockwise   contour is always $2\pi$ corresponding to a vortex of unit angular momentum. It is never $-2\pi$ corresponding to an anti-vortex or multiples of $2\pi$ corresponding to a vortex of multiple angular momentum $(l>1)$. 
}

Next we consider the generation of vortex lattice in the asymmetric case $g_1=1000\ne g_2=900, g_{12}=2000$ briefly illustrated in Fig. \ref{fig2}(c)-(d) and  Fig. \ref{fig8}(a) for $\Omega=0.6$ .  The same for $\Omega=0.7, 0.81$ and 0.9 are displayed in Fig. \ref{fig8} (b)-(d), respectively.
 In this case there is a complete phase separation, but the sizes (extensions) of the components are different.   Also, as before, it is appropriate to display the density of both components in the same 
plot, where it is possible to identify the two components clearly. The maximum density of the condensate decreases with the increase of $\Omega$ and the number of vortices increases with $\Omega$. The filling of the vortices on the components is only approximately of triangular geometry because of the irregular shape of the components of the condensates.

\begin{figure}[!t]

\begin{center}
\includegraphics[trim = 0cm 0.64cm 0cm 0mm, clip,width=\linewidth]{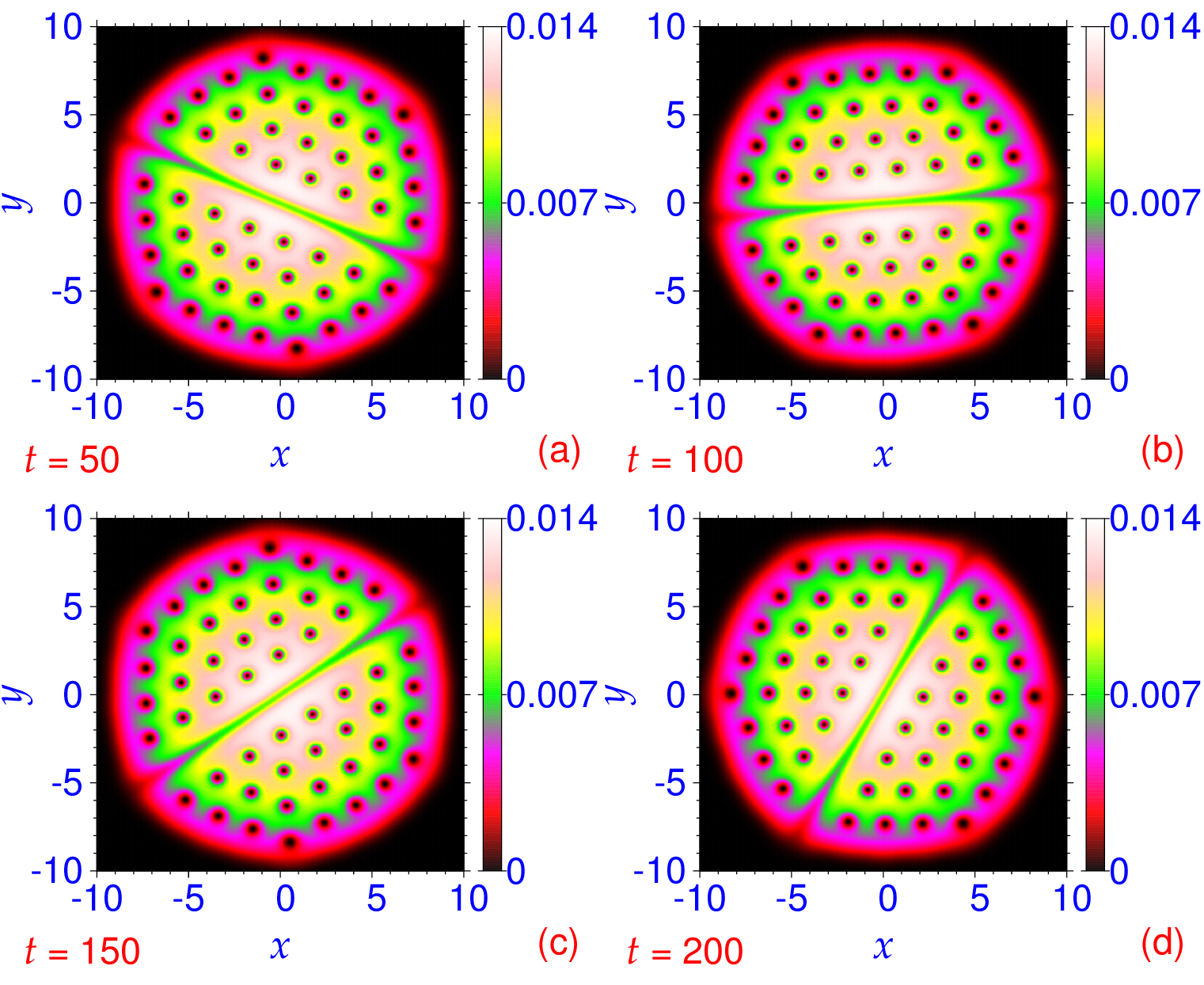} 

\caption{ Dynamical evolution of vortex lattice of a rotating binary BEC,    displayed in Fig.  \ref{fig6}(f), during real-time propagation for 200 units of time
using the corresponding imaginary-time wave function as input, at times (a)
$t = 50$, (b) $t = 100$, (c) $t = 150$, and (d) $t = 200$. During real-time propagation
the angular frequency of rotation $\Omega$  was changed at $t = 0$ from the imaginary-
time value of $\Omega  = 0.88 $ to 0.89. 
}
\label{fig9}
\end{center}

\end{figure}

The dynamical stability of the vortex lattices of the rotating binary BEC is tested next. 
For this purpose we subject the vortex-
lattice state  of the rotating BEC to real-time evolution during a large interval of
time, after slightly changing the angular frequency of rotation $\Omega$ at $t = 0$. The
vortex lattice will be destroyed after some time, if the underlying BEC wave function were dynamically
unstable. We consider real-time propagation of the vortex lattice exhibited in
Fig. \ref{fig6}(f) for $g_1=g_2=1000, g_{12}=2000$ after changing $\Omega$ from 0.88 to 0.89 at $t = 0$. The subsequent evolution
of the vortex lattice is displayed in Fig. \ref{fig9}  at  (a)  $t = 50$, (b) $t = 100$, (c) $t = 150$,
and (d) $t = 200.$  
 The robust nature of the snapshots of vortex lattice during
real-time evolution upon a small perturbation, as exhibited in Fig. \ref{fig9}, demonstrates
the dynamical stability of the vortex lattice in the quasi-2D rotating binary condensate.

\section{Summary and Discussion} 
 
We have studied the generation of spontaneous symmetry-breaking completely   phase-separated circularly-asymmetric vortex lattices in a harmonically-trapped 
repulsive quasi-2D  binary BEC.   In the examples studied in this paper there is no overlap between the component densities of the BEC so that vortex lattices of the two components are formed in different regions of space, which is of great phenomenological interest.  This will facilitate the experimental study of the vortex lattices of the two components. 
Such vortex-lattice structure  is generated when the quantity in Eq. (\ref{eq1}) is much smaller  that unity, e.g., ${g_1g_2}/{g_{12}^2} \lessapprox 0.75$ or so.  For larger values of this ratio, although there could be a phase separation for a  non-rotating binary BEC, overlapping sheet structure \cite{honey,Kasashet} appear for a  rotating binary BEC, viz. Figs. \ref{fig2}(g)-(h). In this study we fixed this ratio as: ${g_1g_2}/{g_{12}^2} \approx 0.5$.  We considered both symmetric and asymmetric cases  of intra-species interaction strengths: $g_1=g_2$ and $g_1\ne g_2$, respectively. In the former case the vortex lattices of the two components lie on fully separated semicircles which are parity conjugates of each other. In the asymmetric case,  although  the phase separation is complete, the shapes of the two components are different.  Of these two, we demonstrated dynamical stability of the symmetric 
vortex lattice by long-time real-time simulation upon a small change in the angular frequency of rotation, viz. Fig. \ref{fig9}. The asymmetric vortex lattice is found to be only weakly stable.     With present experimental know-how these phase-separated vortex lattices can be generated and studied in
a laboratory.

\section*{Acknowledgements}
\noindent

SKA thanks the Funda\c c\~ao de Amparo \`a Pesquisa do
Estado de S\~ao Paulo (Brazil) (Project: 
2016/01343-7) and the Conselho Nacional de Desenvolvimento Cient\'ifico e Tecnol\'ogico (Brazil) (Project:
303280/2014-0) for partial support.


%
\end{document}